\documentclass[11pt]{article}
\usepackage{moriond,epsfig}

\def\be{\begin{equation}}
\def\ee{\end{equation}}
\def\bea{\begin{eqnarray}}
\def\eea{\end{eqnarray}}

\begin{document}
\vspace*{4cm}
\title{\bf Production of $\pi^{\pm}$, K$^{\pm}$, p and $\bar{\rm p}$ in Quark, Antiquark and Gluon Jets}

\author{ Hyejoo Kang }

\address{Department of Physics and Astronomy,Rutgers University,\\
 Piscataway, NJ 08854, USA}

\address{\footnotesize\it Representing the SLD collaboration}

\maketitle\abstracts{We present measurements of identified charged hadron production over a wide momentum range using the SLD Cherenkov Ring Imaging Detector. In addition to studying particle production in flavor-inclusive $Z^{0}$ decays, we compare the production in decays into light, c and b flavor events.
We also examine particle production in gluon jets with that in light quark jets. The jet flavors are selected by using displaced vertex information.}

\section{INTRODUCTION}

The radiation of gluons from the primary quarks in $e^+e^-\rightarrow Z^{0} \rightarrow q\bar{q}$ is principally well understood and caculable through 
perturbative QCD. 
The fragmentation process, by which hadrons are produced from final stage partons, is not fully understood. 
Several phenomenological models of the process~\cite{jetset}~\cite{herwig}~\cite{ucla}, in which the partons radiate gluons that are eventually transformed by different methods into primary hadrons, have been tuned to reproduce data from $e^+e^-$ collisions. To understand the hadronization process better and test these models further, we report measurements of the production of $\pi^{\pm}$, $K^{\pm}$, p and  $\bar{\rm p}$ in inclusive events, $uds$, $c$ and $b$ events and $uds$ quark and gluon jets.

\section{EVENT AND PARTICLE IDENTIFICATION}

We used 55,000 selected~\cite{bfp} hadronic $Z^0$ decays collected by SLD in 1993-98 within the acceptance of the Cherenkov Ring Imaging Detector (CRID). The CRID provides particle identification over a broad momentum range by measuring the opening angle of the cone of Cherenkov light emitted as a charged track passes through liquid and gas radiators. 
To identify a charged track as $\pi^{\pm}$, K$^{\pm}$, p or $\bar{\rm p}$ , a 
likelihood is calculated for each particle type using the number of detected photoelectrons and the number of expected photons and the expected Cherenkov angle. Likelihood ratios can be used to identify a track as a certain type of particle. 
In each momentum bin, identified $\pi$, K, and p are counted, and these numbers are unfolded using the inverse of an identification efficiency matrix, and corrected for track reconstruction efficiency.  The elements of the identification efficiency matrix were determined using a detailed MC simulation of the detector and calibrated with selected $K^{0}_{S}$, $\tau$ decay data and MC.

To separate light($uds$), $c$ and $b$ quarks, we selected secondary vertices using a topological vertex tagging method in each event hemisphere or counted tracks with large impact parameter in event. Since $B$ and $C$ hadrons produce the only secondary vertices and $B$ hadrons have larger vertex mass and more tracks than $C$ hadrons, we can obtain pure and efficient light, $c$ and $b$ quark samples~\cite{bfp}. 

\section{INCLUSIVE HADRONIC FRACTIONS}

\begin{figure}[ht]  
\vskip -0.5cm
\epsfxsize 3.7in
\epsfysize 3.7in
\center\epsffile{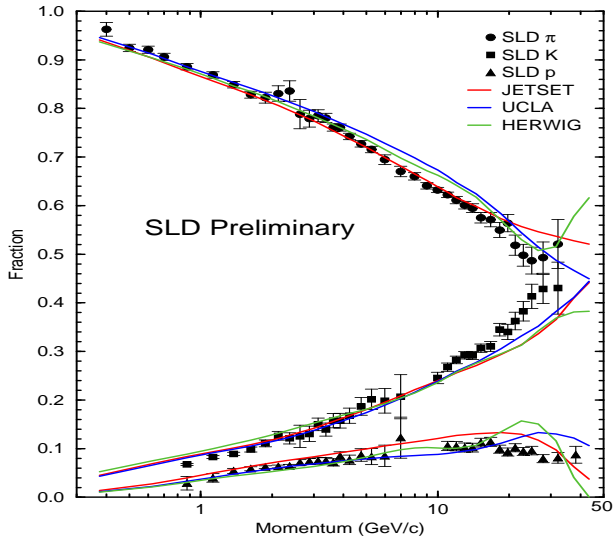} 
\vskip -1.5cm
\caption{\label{fraction}
Comparison of the charged hadron fractions in flavor-inclusive events with the predictions of three fragmentation models}
\end{figure}

The measured charged particle fractions for hadronic $Z^{0}$ decays as a function of momentum are shown in Fig.~\ref{fraction}. Pions are seen to dominate at low momentum and to decline steadily in fraction as momentum increases. The kaon fraction rises gradually to about one-third at high momentum.  The proton fraction rises to a maximum of about one-tenth at momentum $\approx$ 10 Gev/c, then declines. The enhancement of kaons indicates reduced strange supression or that particle production is becoming dominated by leading hadrons.
Also shown in Fig.~\ref{fraction} are the predictions of three fragmentation models with default parameters.  All the models reproduce the shape of
each particle fraction qualitatively.  The HERWIG and UCLA predictions for the pion fraction are high at intermediate momentum. All three predictions for the kaon fraction are too low at high momentum. The JETSET prediction for the proton fraction is too high at all momentum and those of HERWIG and UCLA show structure in the proton fraction at high momentum that is inconsistent with the data.

\section{FLAVOR DEPENDENT ANALYSIS}

\begin{figure}[t]  
\hspace*{15pt}
\hglue 0.1in\parbox{2.8in}{
\epsfxsize 2.8in
\epsfysize 2.65in
\epsffile{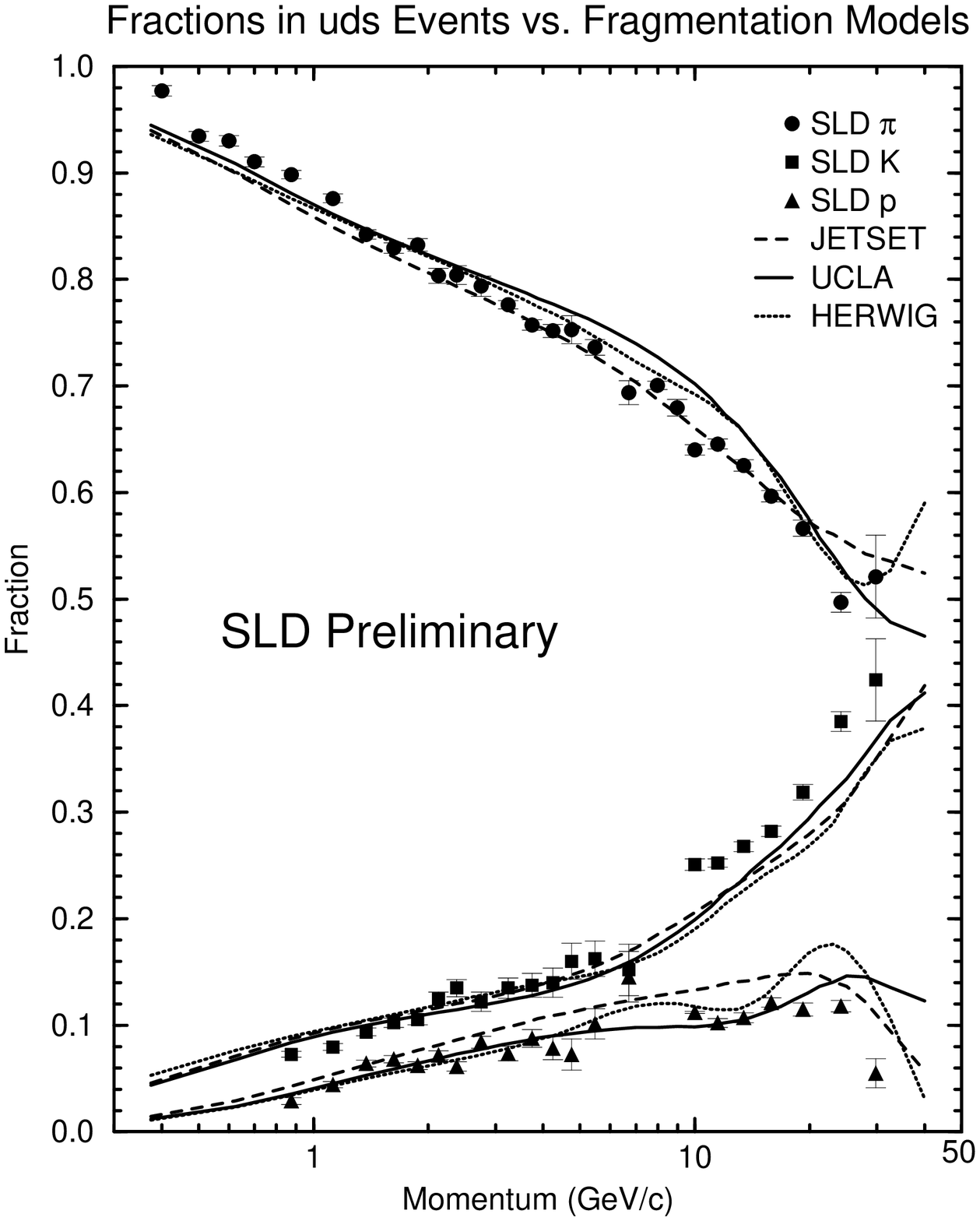}}
\hglue 0.05in\parbox{3.0in}{
\epsfxsize 2.8in
\epsfysize 3.0in
\epsffile{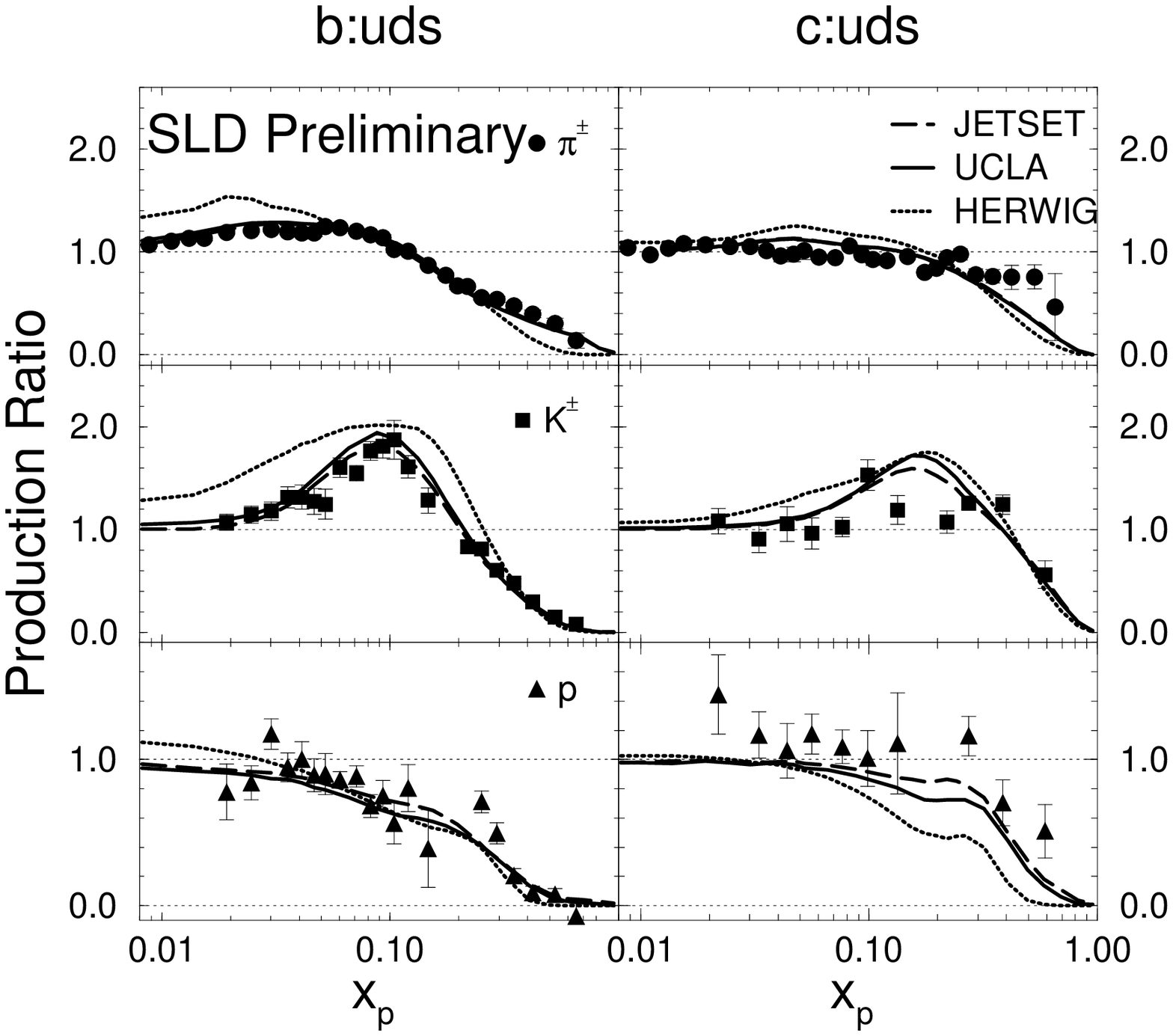}} 
\caption{\label{flvdep}
Comparison of the charged hadron fractions in light-flavor events with the predictions of three fragmentation models(left) and ratios of production rates in $b$- and $c$-flavor events to those in light-flavor events(right)}
\end{figure}

The analysis was repeated separately on the high-purity light, $c$ and $b$ tagged samples and a complete flavor unfolding was done. Fig.~\ref{flvdep} shows the charged hadron fractions in light quark flavor events. Qualitatively there is little difference between these data and those for the inclusive sample, however these are more relevant for comparison with QCD predictions based on the assumption of massless primary quark production.
The same general differences between the predictions of the three models and the data were observed indicating that these deficiencies are in the fragmentation simulation and not simply in the modelling of heavy hadron production and decay.

In Fig.~\ref{flvdep}, the ratios of production in $b$- to light-flavor and $c$- to light-flavor events for the three species are shown. The primary systematic errors on particle identification are cancelled in the ratios, and the predominant errors are statistical.
There is greater production of charged pions in $b$-flavor events at low $x_p=2p/E_{cm}$, with the ratio rising as $x_p$ increses for $0.008 < x_p < 0.03$.
The production of charged kaons is approximately equal in the two samples at $x_p=0.02$, but the relative production in $b$-flavor events increases with $x_p$, peaking at $x_p \approx 0.07$. There is approximately equal production of protons in $b$-flavor and light-flavor events below $x_p=0.15$.   
For $x_p>0.1$, production of all these particle species falls faster with increasing momentum in $b$-flavor events.
These features are consistent with expectations based on the known property
of $b\bar{b}$ events that a large fraction of the event energy is carried by the leading $B$ and $\bar{B}$ hadrons, preventing the production of high momentum hadrons.  The $B$ hadrons decay into a large number of lighter particles which are expected to populate mainly the region $0.02 < x_p < 0.2$.
Similar qualitative features are observed for $c\bar{c}$ events. 
The fragmentation models reproduce these features qualitatively, 
although HERWIG overestimates the pion and kaon ratios by a large factor 
at low $x_p$.

\section{QUARK AND GLUON JET COMPARISON}

\begin{figure}[t]  
\epsfxsize 2.9in
\epsfysize 2.9in
\center\epsffile{ratiogudsq.epsi}
\caption{}
\label{ratio}
\end{figure}

For quark and gluon jets, differences in the inclusive particle production are predicted from QCD and have been observed. These differences are expected to be the same for all identified particle types, except for effects of leading 
particle production and kinematics.
Three-jet events are selected using the Durham algorithm with $y_{cut}$ = 0.005. Jet energies are rescaled using the angles between the jet axes and ordered $E_{1} > E_{2} > E_{3}$. Four different samples of jets are defined by:
\begin{itemlist}

\item Gluon sample: If only one of the lower energy jets has a secondary vertex which passes mass and momentum cuts and jet geometry cuts, the other is tagged as a gluon jet with purity 92\%. 

\item Light mixture: If no secondary vertex and no large impact parameter tracks are found in the event, the two lower energy jets are put into the light mixture sample with $udsg$ purity 94\% and $uds$ quark purity 46\%. 
 
\item $b(c)$ mixture: If the highest energy jet is tagged as $b(c)$,  
the two lower energy jets are included in the $b(c)$ mixture 
with 98(92)\% $b(c)g$ purity.
\end{itemlist}

The fractions analysis is repeated on all four samples. The ratio of each particle's fraction in the tagged gluon jet and light mixture jet samples is shown in Fig.~\ref{ratio} and differs from unity. However the Monte Carlo simulation is consistent with the data and shows that the deviation might be due to a kinematic effects from jet selection bias that must be reduced for future studies.

\section{Conclusion}

Detailed studies of the prodution of $\pi^{\pm}$, K$^{\pm}$, p/$\bar{\rm p}$ are made over a wide momentum range using the pariticle identification with the SLD CRID. The measured particle fractions in flavor-inclusive events are consistent with previous results from SLD and LEP. The fragmentaion models(JETSET, HERWIG and UCLA) describe the data qualitatively but there are still discrepancies for each particle type. 
High purity light($uds$), $c$ and $b$ quark event samples are selected using topological vertex tagging and counting the number of tracks with large impact parameter. Features that are not described by the fragmention models in light quark events are the same in flavor-inclusive events. This indicates that the differences between the data and MC are in the fragmentation not in the modelling of heavy hadron decay.
By comparing the ratio of production in heavy- to light- flavor events, we have shown that there are  additional model discrepancies in heavy hadron decay.
We have seen differences of particle production in gluon and light quark jets, but the differences are reproduced by the MC, indicating kinematic bias from jet selection. 
We have not found any evidence for differences in $\pi^{\pm}$, K$^{\pm}$, p/$\bar{\rm p}$ production within our uncertainty for gluon and light quark, but further studies of the analysis are required.

\end{document}